\documentclass[sigconf,screen]{acmart}

\makeatletter
\def\@ACM@checkaffil{
    \if@ACM@instpresent\else
    \ClassWarningNoLine{\@classname}{No institution present for an affiliation}%
    \fi
    \if@ACM@citypresent\else
    \ClassWarningNoLine{\@classname}{No city present for an affiliation}%
    \fi
    \if@ACM@countrypresent\else
        \ClassWarningNoLine{\@classname}{No country present for an affiliation}%
    \fi
}
\acmYear{2023}\copyrightyear{2023}
\acmConference[ICCPS '23]{ACM/IEEE 14th International Conference on Cyber-Physical Systems (with CPS-IoT Week 2023)}{May 9--12, 2023}{San Antonio, TX, USA}
\acmBooktitle{ACM/IEEE 14th International Conference on Cyber-Physical Systems (with CPS-IoT Week 2023) (ICCPS '23), May 9--12, 2023, San Antonio, TX, USA}
\acmPrice{15.00}
\acmDOI{10.1145/3576841.3589630}
\acmISBN{979-8-4007-0036-1/23/05}

\usepackage[utf8]{inputenc}
\usepackage{enumitem}
\setlist[enumerate]{label*=\arabic*.}
\usepackage{graphicx}
\usepackage{caption}
\usepackage{subcaption}
\usepackage{algorithm,algorithmic}
\usepackage{amsmath}
\usepackage{xfrac}
\usepackage{hyperref}
\usepackage{array}
\usepackage{booktabs}
\usepackage{mathtools}
\usepackage{forest}

\definecolor{mygreenii}{RGB}{91, 198, 98}
\definecolor{mygreeni}{RGB}{148, 188, 151}

\usepackage{titlesec}
\titlespacing{\section}{0pt}{2pt plus 2pt minus 2pt}{2pt}
\titlespacing{\subsection}{4pt}{2pt plus 2pt minus 2pt}{2pt}
\titlespacing{\subsubsection}{4pt}{2pt plus 0pt minus 2pt}{2pt}
\titlespacing{\paragraph}{8pt}{2pt plus 0pt minus 0pt}{2pt}
\newcommand{\urlNewWindow}[1]{\href[pdfnewwindow=true]{#1}{\nolinkurl{#1}}}
\newcommand{\avstack}{\texttt{AVstack}}

\begin{document}
\title{Modular Platform For Collaborative, Distributed Sensor Fusion}\thanks{This work is sponsored by the ONR projects N00014-20-1-2745 and N00014-23-1-2206, AFOSR award FA9550-19-1-0169, NSF CNS-1652544 award and the National AI Institute for Edge Computing Leveraging Next Generation Wireless Networks (CNS-2112562).}

\settopmatter{authorsperrow=5}
\author{R. Spencer Hallyburton}
\affiliation{%
    \institution{Duke University}
}
\author{Nate Zelter}
\affiliation{%
    \institution{Duke University}
}
\author{David Hunt}
\affiliation{%
    \institution{Duke University}
}
\author{Kristen Angell}
\affiliation{%
    \institution{Duke University}
}
\author{Miroslav Pajic}
\affiliation{%
    \institution{Duke University}
}

\begin{abstract}
    Leading autonomous vehicle (AV) platforms and testing infrastructures are, unfortunately, proprietary and closed-source. Thus, it is difficult to evaluate how well safety-critical AVs perform and how safe they \emph{truly} are. Similarly, few platforms exist for much-needed multi-agent analysis. To provide a starting point for analysis of sensor fusion and collaborative \& distributed sensing, we design an accessible, modular sensing platform with \avstack~\cite{2022avstack}. We build collaborative and distributed camera-radar fusion algorithms and demonstrate an evaluation ecosystem of AV datasets, physics-based simulators, and hardware in the physical world. This three-part ecosystem enables testing next-generation configurations that are prohibitively challenging in existing development platforms.
\end{abstract}

\maketitle

\vspace{-1em}
\section{Introduction} \label{sec:1-introduction}

It is unsurprising that many state-of-the-art autonomous vehicles (AV) are proprietary and closed-source: AV manufactures must spend great amounts of money and effort to prove the safety and reliability of their systems. It is important, however, that research and development keep pace with industry so that technology can push forward rapidly \emph{and} so that experts can understand fundamental limitations in performance of safety-critical AV components. Furthermore, emerging concepts in collaborative and distributed autonomy will define the next-generation of autonomy~\cite{2013v2vv2i}, yet there are few existing studies of their efficacy and implementation.

Several platforms have emerged to support open-source AV development. Baidu's Apollo~\cite{BaiduApollo} and Autoware~\cite{Autoware} are preeminent platforms for deployable, real-time AVs. Recently, Pylot~\cite{2021pylot} has allowed for AV component evaluations. Each of these platforms is useful, however, they are highly non-modular. \emph{None of these systems have implemented protocols or algorithms for multi-agent collaboration or distributed computation}, such as vehicle-to-vehicle/infrastructure (V2V, V2I) and edge-computing for low SWaP (size, weight, and power) platforms, respectively.

\begin{figure*}
    \centering
    \includegraphics[width=0.78\textwidth]{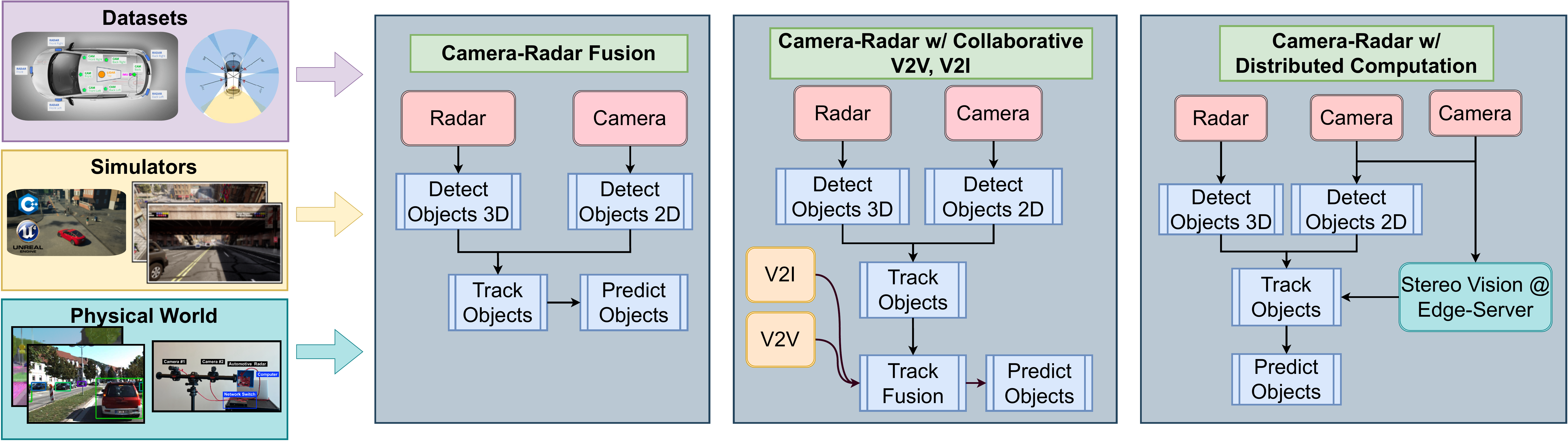}
    \vspace{-1em}
    \caption{Design, implementation, test, and analysis of autonomous vehicles with local, collaborative, and distributed information is set up in a modular configuration with \avstack. Camera-radar fusion (left) is expanded with collaborative vehicle-to-vehicle and vehicle-to-infrastructure sensing (center). With limited-computation, we perform distributed inference at edge servers (right). In each case, we test on datasets, simulators, and in a physical environment (expanded in Fig.~\ref{fig:physical-setup}).}
    \label{fig:platform-full-diagram}
    \vspace{-1em}
\end{figure*}

To promote reusable AV software and to push next-generation evaluations in collaborative and distributed autonomy, we design a modular software platform with \avstack~\cite{2022avstack}. Ours is a unique combination of high-performing algorithms with a flexible and reconfigurable software architecture for diverse and rapid prototyping. This instantiation is based around a central camera-radar fusion paradigm. We build two next-generation designs with (1) collaborative V2V, V2I tracking, and (2) distributed computation for intensive algorithms. These designs are illustrated in Fig.~\ref{fig:platform-full-diagram}. 

We build a testing ecosystem from the bottom up: (i) components are validated on datasets from real-world captures, (ii) AV designs are deployed onto simulators in custom configurations, (iii) AVs run in near-real-time while ingesting data from a physical sensing rig. This three-part testing ecosystem ensures that components receive adequate objective evaluation early in the AV lifecycle. It also allows testing unique sensor fusion, collaborative, and distributed configurations that are prohibitively challenging in existing development platforms. Our software including algorithms and testing ecosystems has been released open-source~\cite{AVStack}.


\section{Modular Software Design With AVstack, ZMQ} \label{sec:2-software}

Our software is built with three primary goals in mind:

\begin{enumerate}[start=1,label={(\bfseries \roman*)}]
    \item Multi-Source Testbed (i.e.,~datasets, simulators, hardware)
    \item Multi-Sensor, Multi-Platform (i.e.,~V2V, V2I) Integration 
    \item Algorithm Reconfigurability
\end{enumerate}

To achieve these goals, we leverage existing application-layer and transport-layer libraries \avstack~\cite{2022avstack} and ZMQ~\cite{2013zmq}. \avstack\ enables implementation of processes such as perception, tracking, and sensor~fusion and maintaining data-source agnostic interfaces. ZMQ~\cite{2013zmq} supports management of communication channels between threads, processes, sensors, and platforms. 

\subsection{Multi-Source Testbed}

\paragraph{Data Types.} \avstack\ is data-source agnostic. We leverage this flexibility to establish a multi-stage AV test ecosystem using: (i) KITTI~\cite{2013kittidataset} and nuScenes~\cite{2020nuscenesdataset} datasets, (ii) CARLA simulator~\cite{2017carla}, and (iii) physical platforms with near-real-time requirements.

\paragraph{Network Interfaces.} The ego vehicle performs many inference tasks (e.g.,~object detection) natively. Over direct connections, we integrate: (i) Gigabit Ethernet (GigE) cameras, (ii) Camera Serial Interface (CSI) cameras, and (iii) Serial Bus (USB) radar. Over a wireless interface, the ego can send tasks to edge servers for distributed computation. Similarly, it can receive sensor data or scene information from neighboring entities (e.g.,~V2V, V2I). Due to the plurality and heterogeneity of these connections, a major component of the software design is implementing flexible communication channels.

\paragraph{Evaluation.} At each stage of testing - datasets, simulators, hardware - we benchmark implementations with objective metrics at the perception, tracking, and motion-prediction levels. This helps to validate the selection of algorithms and our implementations.

\subsection{Multi-Sensor, Multi-Platform}

To encourage adoption of multi-sensor, multi-agent AVs, we test multiple implementations of sensor fusion architectures.

\paragraph{Camera-Radar (C+R) Fusion.} A single camera feeds a 2D bounding box detector. These detections are fused with 3D point-wise detections from radar. This simple fusion scheme is popular among self-driving manufacturers~\cite{CommaAI}. See Fig.~\ref{fig:platform-full-diagram}, left. 

\paragraph{C+R Fusion with Collaborative V2V/I (C+R-CoVI).} 
Collaborative vehicles and infrastructure provide high-level information about the scene/objects that is fused with local tracks in a dedicated fusion module. See Fig.~\ref{fig:platform-full-diagram}, middle.

\paragraph{C+R Fusion with Distributed Computation (C+R-Dist).} 
In parallel, images are sent to an edge-server where computation-intensive stereo image-processing is performed. This system represents a case where computation is limited on the ego. See Fig.~\ref{fig:platform-full-diagram}, right.

\subsection{Algorithm Reconfigurability}

With \avstack, we can design a modular software base that can be reconfigured with minimal overhead to meet each AV design's needs. C+R, C+R-CoVI, and C+R-Dist are based on reusable perception and tracking components. 
\section{Testing Platforms} \label{sec:3-Platform}


We use \avstack\ and ZMQ to set up three testbeds. First, we implement a data-replay system for running our AVs on implementations from longitudinal datasets. The second platform uses CARLA, an AV simulator, to stand up a multi-agent simulation. Finally, we design and build a physical platform with real sensors and test in an urban environment.

\paragraph{Datasets and Simulators.} Evaluations begin on AV datasets from real-world driving captures. The software platform runs on longitudinal scenes from KITTI~\cite{2013kittidataset} and nuScenes~\cite{2020nuscenesdataset} with objective metrics at each step in the AV pipeline. We then move to the physics-based simulator CARLA~\cite{2017carla} for closed-loop testing in dynamic environments with custom sensor and agent configurations.

\begin{figure}[t!]
    \centering
    {%
    \setlength{\fboxsep}{2pt}%
    \setlength{\fboxrule}{2pt}%
    \fbox{\includegraphics[width=0.68\linewidth]{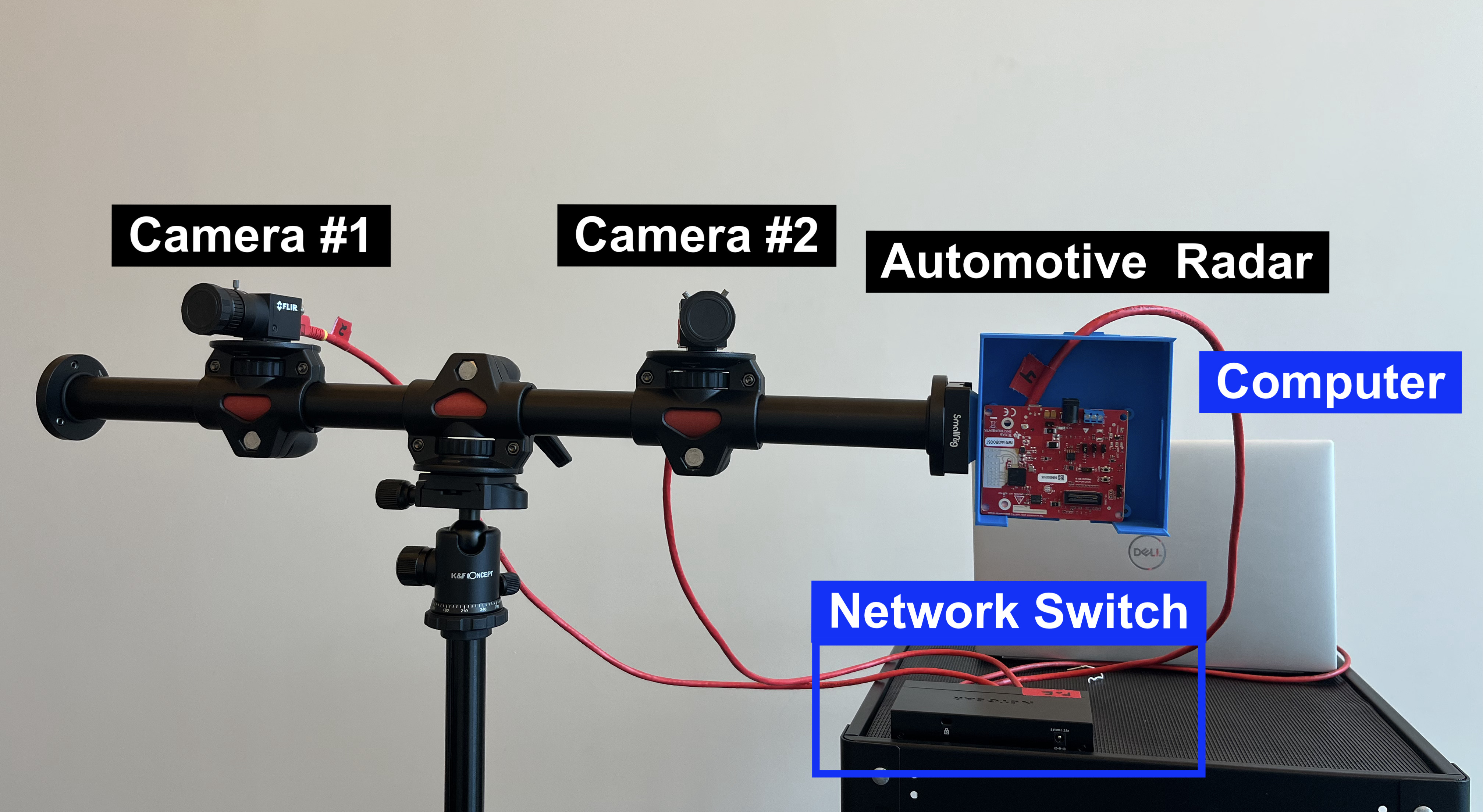}}%
    }
    \vspace{-1em}
    \caption{Physical sensing setup. Cameras and radar connected via wired interface. Computer hosts algorithms and wireless interface for V2V, V2I, and edge-computing.}
    \label{fig:physical-setup}
    \vspace{-1.2em}
\end{figure}

\paragraph{Physical Sensing System.} We build a sensing system with a diverse array of commercially-available components. We use two FLIR BlackFly S cameras over a gigabit ethernet connection together with infrastructure systems built with the RaspberryPi high-quality camera to form our vision subsystem. We build a multi-modal fusion system with the addition of a Texas Instruments IWR-1443 automotive radar. The physical sensing system is illustrated in Figure~\ref{fig:physical-setup} and will be presented during demonstration.

\section{Conclusion}

We built a software ecosystem for modular testing of camera-radar fusion with collaborative and distributed computation. We illustrated that such implementations can be tested on datasets, simulators, and in real-world sensing scenarios with minimal software overhead using \avstack\ and ZMQ. Our platform is a starting point for rapid AV prototyping, test, and analysis with an emphasis on sensor fusion, collaborative sensing, and distributed computation.


\bibliographystyle{ACM-references}
\bibliography{references}

\end{document}